\documentstyle[prc,preprint,tighten,aps]{revtex}
\begin{document}
\draft
\title{Coulomb Breakup Mechanism of Neutron-Halo Nuclei
in a Time-Dependent Method}
\author{T. Kido$^1$, K. Yabana$^2$, and Y. Suzuki$^2$
\
\
\\$^1$Graduate School of Science and Technology,
Niigata University, Niigata 950-21, Japan
\\$^2$Department of Physics, Niigata University, Niigata 950-21, Japan}
\maketitle

\begin{abstract}
The mechanism of the Coulomb breakup reactions of the nuclei
with neutron-halo structure is investigated in detail. A time-dependent 
Schr\"odinger equation for the halo neutron is numerically solved by 
treating the Coulomb field of a target as an external field.
The momentum distribution and the post-acceleration
effect of the final fragments  are discussed in a fully quantum 
mechanical way to clarify the limitation of the  
intuitive picture based on the classical mechanics.
The theory is applied to the Coulomb breakup reaction of $^{11}$Be + 
$^{208}$Pb. The breakup mechanism is found to be different between the
channels of $j^{\pi}=\frac{1}{2}^{-}$ and $\frac{3}{2}^{-}$, reflecting 
the underlying structure of $^{11}$Be. The calculated result reproduces 
the energy spectrum of the breakup fragments reasonably well, but   
explains only about a half of the observed longitudinal momentum 
difference.
\end{abstract}


\section{Introduction}
\quad

The neutron-halo structure has been observed systematically \cite{Hansen} 
in light neutron-rich nuclei near the neutron drip-line. 
A large Coulomb breakup cross section has been observed for the 
neutron-halo nucleus in reactions on a heavy target nucleus 
\cite{Kobayashi}, 
which indicates that a significant amount of $E1$ strength exists at low 
excitation energy region of the neutron-halo nucleus \cite{Suzuki}. Since 
stable nuclei do not have such a strong $E1$ distribution at low excitation 
energy, this unusual feature of the neutron-halo nucleus has attracted much 
attention. The origin of the strong $E1$ distribution is still 
controversial: 
It may be a resonant character due to the vibration of the halo neutron 
against a core nucleus \cite{Ikeda}. Or it may be understood from an 
analogy of the strong $E1$ transitions known in 
$^9$Be, $^{11}$Be, and $^{13}$C where a single nucleon is weakly coupled 
to a core nucleus \cite{Millener}.
A recent argument suggests that the low-lying $E1$ strength of
the light halo nuclei is not considered the vibrational state 
\cite{Kurasawa}.

Recent experiments of the Coulomb breakup reactions, $^{11}$Li 
+ $^{208}$Pb \cite{Ieki} and $^{11}$Be +$^{208}$Pb \cite{Nakamura},
have observed  
a significant longitudinal momentum difference 
between the halo neutron(s) and the core nucleus. 
The momentum difference has been explained in terms of the Coulomb 
post-acceleration effect by assuming a direct breakup mechanism. 
In this mechanism the breakup is assumed to
occur instantaneously at the closest approach point between the projectile 
and target nuclei. After the breakup of 
the projectile nucleus, the target Coulomb field accelerates only the 
core nucleus, and causes 
the momentum difference between the neutron and the core nucleus.
If the breakup proceeds through a resonant state of the projectile nucleus, 
the core nucleus and
the halo neutron moves together during its lifetime and  
the post-acceleration effect should become small. The 
observation of the longitudinal velocity difference  is thus recognized as 
a direct evidence for the non-resonant character of the $E1$ strength at
low excitation energy.

The post-acceleration effect is not explained in the lowest order 
perturbation
treatment of the Coulomb excitation. The evaluation of higher order effects 
is not easy because the final 
states involve continuum states. Several
theoretical approaches have been proposed to understand the 
post-acceleration effect, including a classical
treatment of the breakup reaction \cite{Baur}, a distorted-wave
Born approach \cite{Shyam},
a simplified treatment of the higher order perturbations \cite{Typel}, and 
a coupled-channel approach with discretized continuum states \cite{Canto}.
Contrary to these approaches, some groups have investigated the time
evolution of the projectile nucleus by solving a time-dependent 
Schr\"odinger equation on mesh points of space and time variables 
\cite{Bertsch,Kido}. 

In our previous paper \cite{Kido} we studied the Coulomb breakup of 
$^{11}$Be and found 
large transverse and small longitudinal momentum differences between the 
neutron and the $^{10}$Be nucleus. The result was understood in the 
picture of 
free-particle breakup mechanism which is in contrast to the direct breakup 
mechanism. However, the reproduction of the 
experimental momentum difference remained to be an open problem. In this 
paper we extend the previous calculation to a more realistic case by 
including the spin-orbit interaction between the neutron and the core 
nucleus. A full three-dimensional dynamical calculation has also been 
done in [13b] assuming a simple internal Hamiltonian. No investigation has,  
however, so far been performed to clarify the roles 
of the $ls$ potential and the level structure of excited states. We first 
analyze the mechanism of the Coulomb breakup 
quantum-mechanically for various cases of the neutron-core Hamiltonian.
The results are discussed in comparison with the intuitive arguments
based on the classical mechanics. The usefulness and  
limitation of the classical arguments is made clear.
We then analyze the Coulomb breakup reaction of $^{11}$Be + $^{208}$Pb 
\cite{Nakamura}
with a realistic choice of the potential between the 
halo neutron and $^{10}$Be. We show that the breakup mechanism is sensitive 
to the structure of the excited states of the projectile and that 
the inclusion  
of the spin-orbit interaction is very important to reproduce the observed  
features quantitatively. 

In section 2 we formulate the quantum-mechanical treatment of the Coulomb 
breakup reaction using the time-dependent Schr\"odinger equation. The method 
of calculating the time evolution of the wave function is briefly 
explained. In 
section 3 the theory is applied to various cases of the halo nucleus to 
discuss the limitation of the classical arguments for the Coulomb breakup 
reaction and to reveal the characteristics of the quantum aspect. The 
Coulomb breakup of $^{11}$Be on a $^{208}$Pb target is analyzed in section 
4. A brief summary is given in section 5.  

\section{Formulation}
\quad

We consider the Coulomb breakup reaction of the nucleus with 
neutron-halo structure. The projectile of the halo nucleus is assumed to 
consist of a single neutron and a core nucleus. The core nucleus 
is treated as a structureless 
particle and binds the neutron weakly by an appropriate potential.
We describe the reaction in the projectile rest frame where the 
center-of-mass of the 
projectile is put at the origin of the coordinate. 
The time development of the  wave function, $\Psi({\bf r},t)$, of the 
relative motion between the neutron and the core nucleus 
is described by the following time-dependent Schr\"odinger equation,
\begin{equation}
i\hbar {\partial \over \partial t} \Psi ({\bf r},t)
= \bigl\{ H({\bf r}) + V_{\rm ext}({\bf r},t) \bigr\} \Psi ({\bf r},t),
\end{equation}
where $H({\bf r})=-\frac{\hbar^2}{2\mu}\nabla^2+V({\bf r})$ is the 
internal Hamiltonian describing the relative
motion between the halo neutron and the core nucleus. 
We assume that the projectile moves along a straight line trajectory  
with a constant velocity. The target nucleus exerts a Coulomb potential  
$V_{\rm ext}({\bf r},t)$ on the projectile. The potential 
is treated as a time-dependent external field in the projectile rest frame,
\begin{equation}
V_{\rm ext}({\bf r},t) = {Z_C Z_T e^2 \over
| {m_n \over m_n + M_C}{\bf r}+{\bf b}+{\bf v}t | }
- {Z_C Z_T e^2 \over |{\bf b}+{\bf v}t| }.
\end{equation}
Here ${\bf b}$ is the impact parameter which specifies the straight line 
trajectory, ${\bf v}$ is the incident velocity of the projectile, and  
$Z_T$ and $Z_C$ are the charge numbers of the 
target and core nucleus, respectively. The masses of the neutron 
and the core nucleus are denoted by $m_n$ and $M_C$, respectively.

The wave function is expanded in partial waves as 
\begin{equation}
\Psi({\bf r},t)=
\sum_{lm}\frac{u_{lm}(r,t)}{r}Y_{lm}(\hat{\bf r}).
\end{equation}
When the internal Hamiltonian $H({\bf r})$ includes the spin-orbit 
interaction, it is convenient to couple the spin of the neutron 
with the relative orbital angular momentum $l$ to the total angular 
momentum $j$. Since the generalization to such a case 
is straightforward, we develop the formulation by assuming the wave 
function of Eq.(3).

We descretize the time variable in a step $\Delta t$, and represent 
the wave function of $n$-th time step as $\Psi^{(n)}({\bf r})$. The wave 
function of $(n+1)$-th time step  is calculated by the following formula,
\begin{equation}
{\displaystyle
\Psi^{(n+1)}({\bf r}) \simeq 
e^{-iH({\bf r}) \Delta t/\hbar}
e^{-iV_{\rm ext}({\bf r},t) \Delta t/\hbar}\Psi^{(n)}({\bf r}).}
\end{equation}
The radial part of the wave function of $n$-th step is denoted as 
$u^{(n)}_{lm}(r)$.
The time development is then achieved by two successive procedures.
First, the development due to the external field $V_{\rm ext}({\bf r},t)$ 
is approximated by 
\begin{equation}
u^{(n+\frac{1}{2})}_{lm}(r) =
u^{(n)}_{lm}(r) - i \Delta t/ \hbar \sum_{l'm'}
\langle lm| V_{\rm ext}({\bf r},t) |l'm'\rangle u^{(n)}_{l'm'}(r).
\end{equation}
The evaluation of the matrix element in Eq. (5) is done by expanding the 
external field into multipoles. 
Next the time development by the internal Hamiltonian 
is performed separately for each
angular momentum channel, $lm$, by using the following approximation  
\begin{equation}
u^{(n+1)}_{lm}(r) =
\frac{ 1 - i \frac{\Delta t}{2\hbar} h_l(r)}
     { 1 + i \frac{\Delta t}{2\hbar} h_l(r)}
u^{(n+\frac{1}{2})}_{lm}(r),
\end{equation}
with 
\begin{equation}
h_l(r) = -\frac{\hbar^2}{2\mu} \frac{d^2}{d r^2}
+ \frac{\hbar^2 l(l+1)}{2\mu r^2} + V_l(r).
\end{equation}
To obtain $u^{(n+1)}_{lm}(r)$ from $u^{(n+\frac{1}{2})}_{lm}(r)$, we 
discretize the radius variable $r$ on mesh points of an equal 
spacing and employ the Crank-Nicolson formula \cite{Koonin} using 
three-point formula for the second order differential operator. 

The time evolution is calculated according to the above prescription from 
an initial wave function, $\Psi({\bf r},t=-\infty)=\phi_0({\bf r})$, which 
is the ground-state wave function of the internal Hamiltonian. 
The breakup component of the wave function is
obtained by eliminating all the bound state components of the Hamiltonian,
\begin{equation}
{\displaystyle
|\Psi^{\rm BU}_{\bf b}({\bf r},t) \rangle
=\Bigl(1-\sum_{i \in {\rm bound}} |\phi_i \rangle \langle \phi_i|\  \Bigr)\  
|\Psi({\bf r},t) \rangle.
}
\end{equation}
Here the subscript ${\bf b}$  is kept to stress that the time evolution 
of the wave function is calculated for each impact parameter of the 
external field.

The momentum distribution of the relative motion between the neutron and 
the core nucleus after the breakup is obtained by 
\begin{equation}
{\displaystyle
{dP_{\rm BU}({\bf b},{\bf k}) \over d{\bf k} }
=\lim_{t \rightarrow \infty}| \langle{\bf k} |\Psi^{\rm BU}_{\bf b}
({\bf r},t) \rangle|^2,}
\end{equation}
where $|{\bf k}\rangle$ represents the plane wave state.
The integration of Eq. (9) over ${\bf k}$ yields the Coulomb breakup
probability, $P_{\rm BU}(b)=\lim_{t \rightarrow \infty}
\langle \Psi^{\rm BU}_{\bf b}({\bf r},t) | \Psi^{\rm BU}_{\bf b}({\bf r},t) 
\rangle$, for a given $\bf b$. Integrating 
$P_{\rm BU}(b)$ over the 
impact parameter vector yields the total Coulomb breakup cross section. 

The breakup cross section can be expressed as a function of the
relative motion energy $E$ between the neutron and the core nucleus 
as follows,
\begin{equation}
{d\sigma_{\rm BU} \over d{E}} =  2\pi
\int_{b_{\rm min}}^{\infty} d b \ {b}
\ \int d{\bf k}  \ \delta \Bigl( E_{\bf k} - E \Bigr)
{d {P}_{\rm BU}({\bf b},{\bf k}) \over d{\bf k} }, 
\end{equation}
where $E_{\bf k}={\hbar^2 k^2 \over 2 \mu}$. The convergence of the 
integral in Eq. (10) is very slow with respect to 
the impact parameter. 
Furthermore the calculation of the relative momentum distribution requires 
a long time step  for large impact parameters. To circumvent this 
difficulty, we divide the integration interval of $b$ to two parts, 
$[b_{\rm min}, b_s]$ and $[b_s,\infty]$, and in the latter interval   
employ the first order perturbation theory to calculate the energy 
distribution. The value of $b_s$ is chosen in such a way that  
the first order perturbation theory (PT) and the time-dependent 
calculation (TD) give approximately the same energy distribution at 
$b=b_s$. Equation (10) is then recast to  
\begin{equation}
{d\sigma_{\rm BU} \over d{E}} =
{d\sigma_{\rm BU}^{\rm (TD)} \over d{E}}
+{d\sigma_{\rm BU}^{\rm (PT)} \over d{E}}.
\end{equation}
The second term of Eq. (11) is expressed in a closed form by using the 
perturbation theory. For this aim we use the breakup probability 
distribution which is obtained in the first order perturbation theory,  
\begin{equation}
{d{P}_{\rm BU}^{\rm (PT)}(b) \over d{E}}
={16 \pi \over 9}{{Z_T}^2 e^2 \over (\hbar v)^2}
 ({\xi \over b})^2 [K_0^2(\xi)+K_1^2(\xi)]{d{B}(E1) \over d{E}},
\end{equation}
where $K_0$ and $K_1$ are the modified Bessel functions and $\xi=
{bE \over \hbar v}$. 
By integrating Eq. (12) over the impact parameter in the interval 
$[b_s,\infty]$, we obtain
\begin{equation}
{d \sigma_{\rm BU}^{\rm (PT)} \over d{E}}
={16 \pi \over 9}{{Z_T}^2 e^2 \over (\hbar v)^2}
 2\pi \xi_s K_0(\xi_s)K_1(\xi_s){d{B}(E1) \over d{E}}, 
\end{equation}
where $\xi_s={b_s E \over \hbar v}$. 

The $B(E1)$ strength function in Eqs. (12) and (13) is defined by
\begin{eqnarray}
{d{B}(E1) \over d{E}}
&\equiv&\sum_{Mi}|\langle \phi_i|D_{1M}|\phi_0 \rangle|^2\delta(E_i-E)
 +\sum_M \int d{\bf k} |\langle \phi_{\bf k}|D_{1M}|\phi_0 \rangle|^2
\delta(E_{\bf k}-E)
\nonumber\\
&=&\sum_M \langle \phi_0|D_{1M}^{\dag}\delta(\hat{H}-E)D_{1M}|\phi_0\rangle
\nonumber\\
&=-&{1 \over \pi}
\sum_M {\rm Im}\langle \phi_0|D_{1M}^{\dag}{1 \over E-H+i\varepsilon}
D_{1M}|\phi_0\rangle\\
\nonumber
\end{eqnarray}
with the dipole operator $D_{1M}$ defined by
\begin{equation}
D_{1M}=-\frac{Z_C}{A_C+1}erY_{1M}(\hat{\bf r}), 
\end{equation}
where $A_C$ is the  mass number of the core nucleus. The wave 
functions $\phi_i$ and $\phi_{\bf k}$ of Eq. (14) are the bound excited 
states and  
continuum states of the Hamiltonian $H({\bf r})$ and $E_i$ and $E_{\bf k}$ 
are the corresponding eigenvalues, respectively. To calculate 
the $B(E1)$ strength function we rewrite Eq. (14) in a time-dependent form, 
\begin{eqnarray}
{d{B}(E1) \over d{E}}
&=&{3 \over \pi \hbar}{\rm Re}\langle \phi_0|D_{10}^{\dag} \int_0^{\infty}dt
e^{{i \over \hbar}(E+i\varepsilon)t} e^{-{i \over \hbar}Ht}D_{10}|
\phi_0\rangle \nonumber \\
&=& {3 \over \pi \hbar} {\rm Re} \int^{\infty}_0 dt
e^{{i \over \hbar}(E+i\epsilon)t} \langle \phi_0 | D^{\dag}_{10} |
\psi(t) \rangle, 
\end{eqnarray}
where the ground state is assumed to have $l=0$. The wave function 
$\psi(t)=e^{-{i \over \hbar}Ht}D_{10}\phi_0$ satisfies the time-dependent 
Schr\"odinger equation
with the Hamiltonian $H({\bf r})$ and its initial wave function $\psi(0)$ 
is equal to $D_{10} \phi_0$. By solving the time-dependent
Schr\"odinger equation without an external field, it is possible to obtain 
the wave function $\psi(t)$ and then calculate the $B(E1)$ strength 
function according to Eq. (16). It is straightforward to generalize 
Eq. (16) to the case where the spin-orbit potential is included.

\section{Quantum-Mechanical Analysis of Coulomb Breakup Process}
\quad

To elucidate the breakup mechanism of a neutron-halo nucleus, we
investigate the breakup process for various choices of the internal 
Hamiltonian $H({\bf r})$.
Before showing calculated results, we first discuss typical intuitive 
pictures for the breakup mechanism based on the classical mechanics. 
They include:
\begin{enumerate}
\item Direct breakup mechanism
\item Resonant breakup mechanism
\item Free-particle breakup mechanism
\end{enumerate}

\noindent
In the direct breakup mechanism, the neutron is assumed to be removed
suddenly from the core nucleus when the projectile nucleus approaches the 
point closest to the target nucleus.  
After the breakup occurs, only the core nucleus is 
accelerated by the target Coulomb field. The relative momentum between 
the neutron and the core nucleus is finite both 
for the longitudinal and transverse directions.
When the projectile nucleus has such a resonance state that can be excited
from the ground state by the target Coulomb field, the resonant breakup 
process may become important and the breakup may 
proceed dominantly by way of the resonance. If the lifetime
of the resonance is long enough, the neutron and the core nucleus would 
move together for a long time before the breakup occurs. In such an 
extreme case there is no difference in the relative momentum 
between the neutron and the core nucleus.
In the last case of free-particle breakup mechanism we consider the 
limiting case where the binding energy of 
the neutron is extremely small. Then the neutron and the core nucleus
behave independently during the reaction. The core nucleus moves along
the Rutherford trajectory, while the halo neutron, receiving no Coulomb 
force from the target, moves on the straight line.
In the end the core nucleus receives the momentum only in the transverse 
direction. The momentum difference between the neutron and the core nucleus 
arises only in the transverse direction.
Table I summarizes the momentum differences for the three cases discussed 
above.

It is not clear how well the above classical arguments are quantitatively 
correct, though they are  easily accepted intuitively.  
In what follows we show results of the quantum-mechanical calculations 
for various cases of the neutron-core interactions for which the 
above-mentioned reaction mechanisms are expected to be manifest.

We choose various parameters for the study 
of the breakup reaction of $^{11}$Be 
on a $^{208}$Pb nucleus performed at the incident energy of 72 MeV/nucleon 
 \cite{Nakamura}. 
Hence the projectile nucleus of $^{11}$Be has the incident velocity of  
$v/c = 0.37$. By choosing the reaction plane to be $x$-$z$ plane, the 
target nucleus moves on the straight line ${\bf R}_T(t) = (b,0,-vt)$ in the 
projectile rest frame. The impact parameter $b$ is fixed to $b=12$ fm in 
the present section. The radius variable $r$ is taken up to 800 fm and it 
is descretized with the mesh size of 
$\Delta r = 0.4 $ fm. The time step $\Delta t/\hbar = 0.01$ MeV$^{-1}$ is 
used for calculating the time development of the  wave function.  
The Schr\"odinger equation is solved for the time interval of 
$-10 \le t/\hbar \le 10$ MeV$^{-1}$. At the initial and
final stages of the calculation, the target nucleus is apart from 
the projectile's center-of-mass coordinate by about 750 fm in the 
longitudinal direction.

We assume that the $\frac{1}{2}^+$ ground state of the $^{11}$Be
nucleus is described with a single neutron-halo structure 
in the $1s$ orbit around the inert $^{10}$Be core. The potential 
between the
neutron and the core nucleus is taken to be a spherical Woods-Saxon 
potential.  
The spin-orbit interaction is turned off in the calculation presented 
in this section. The radius and diffuseness parameters of the potential are 
fixed to $R = 2.67$ fm and $a=0.6$ fm, whereas the depth of the potential 
is treated as a variable parameter to investigate the breakup mechanism. 
The partial waves up to $l=4$ are included in the expansion of Eq. (3). 
The contribution of higher partial waves is found to be negligible. 
The target Coulomb field of Eq. (2) is expanded in multipoles around the 
projectile's center-of-mass, and the dipole and quadrupole multipoles are 
included in the calculation. The dipole field plays a dominant role in the 
present system. The contribution of the quadrupole field is found to be 
small. 

The average value of the relative momentum between the neutron and 
the core nucleus is calculated by the following formula
\begin{equation}
\langle {\bf k}\rangle =\lim_{t \rightarrow \infty} 
\frac{ \langle \Psi^{\rm BU}_{\bf b}({\bf r},t)|-i {\nabla} |
\Psi^{\rm BU}_{\bf b}({\bf r},t) \rangle}
{\langle \Psi^{\rm BU}_{\bf b}({\bf r},t)|\Psi^{\rm BU}_{\bf b}({\bf r},t) 
\rangle}. 
\end{equation}
Similarly the average values of the longitudinal and transverse momentum 
differences, $\langle k_{\parallel}\rangle$ and $\langle k_{\perp}\rangle$, 
are defined by the breakup component 
$\Psi^{\rm BU}_{\bf b}({\bf r},t)$. We use the convention that the 
longitudinal 
and transverse directions indicate the $z$ and $x$ directions, respectively. 

\subsection{Dependence on the neutron binding energy}
\par\indent
To discuss the validity of the classical arguments summarized in Table I, 
we first show the result of calculation obtained by changing the 
neutron-core potential depth $V_0$ of $l=0$ channel. The value of $V_0$ is 
varied to understand the dependence of the breakup reaction mechanism on 
the binding energy of the halo neutron of the $1s$ orbit. The potential 
depth of other channels, denoted $V_p$, is set the same as $V_0$. 
The $0p$ orbit is always bound below the $1s$ orbit, and 
there is no resonant nor bound excited state which can be excited from 
the ground state by the dipole field. The halo nucleus considered in this 
subsection is thus a very simple system, and excited to the continuum 
directly by the Coulomb field.

Figure 1 displays the magnitude of $\langle k_{\parallel}\rangle$
and $\langle k_{\perp} \rangle$ values as a function of $V_0$. 
The lower part of Fig. 1 shows the binding energy
of the $1s$ orbit. As the binding becomes weaker, the longitudinal momentum  
difference decreases and approaches zero in the vanishing binding energy, 
while the transverse momentum difference increases slightly. This result 
is consistent with the free-particle breakup picture. However, the 
transverse relative momentum is always smaller than the value, 
$2 k_c = 0.092$ fm$^{-1}$, which is expected from the classical argument.  
On the other hand, in the strong binding case, both of the longitudinal and 
transverse momentum differences are finite and close to the value of 
$k_c=0.046$ fm$^{-1}$, which is just expected from the direct 
breakup picture. These calculations including both weak and strong binding 
cases indicate that the breakup mechanism 
of the neutron-halo nucleus, whose binding energy is typically less 
than 1 MeV, proceeds between the two mechanisms of the direct and 
free-particle breakup. The free-particle picture 
becomes more suitable with the decreasing binding energy of the halo neutron.

We show in Fig. 2 the $B(E1)$ strength  as a function of the
neutron-core relative energy. The value of $\varepsilon$ of Eq. (16) is 
set 0.01 MeV. The $B(E1)$ strength function is closely related to the 
energy spectrum of 
the breakup cross section (see Eq.(13)). The non-perturbative
effect gives only a small effect on the energy spectrum of the cross 
section [14,13b].
The $B(E1)$ strength is shown for two cases of the ground-state energy: 
$E=-2.00$ MeV ($V_0 = -65.2$ MeV), and $E=-0.503$ MeV
($V_0 = -58.3$ MeV). The latter value is chosen to fit the empirical 
neutron separation energy of $^{11}$Be.
No resonance exists in $p$-wave for both cases. Therefore the peak in 
the $B(E1)$ strength which appears at low excitation energy in case of 
$E=-0.503$ MeV has nothing to do with any resonant character. We see that 
the energy spectrum of the $B(E1)$ strength is very sensitive 
to the neutron binding energy. Since the breakup
probability is sensitive to the $E1$ strength at low excitation energy, the
breakup cross section is also sensitive to the neutron binding energy. 

\subsection{Dependence on the level structure of excited states}
\par\indent
Here we investigate how the breakup mechanism depends on the structure 
of the excited states of the halo nucleus. For this purpose 
we employ the angular momentum ($l$-) dependent neutron-core potential. 
The potential depth of $l=0$ channel is fixed to $V_0 = -58.3$ MeV in 
order to fit the energy of the $1s$ orbit to $E_0 = -0.503$ MeV, the 
ground-state energy of $^{11}$Be from the $n+^{10}$Be threshold. The value 
of $V_p$ is now treated as a variable parameter. 
As shown in the lower part of Fig. 3, there is a bound $0p$ state  
when $V_p \le -31.3$ MeV. In case of 
$ -$33.2 MeV $< V_p < -31.3 $ MeV, the bound $0p$ state appears 
between the $1s$ ground state and the neutron threshold. This $0p$ state 
is excited by the dipole field in the breakup reaction. 
In case of $-$31.3 MeV $\le V_p \le -30$ MeV a $p$-wave resonance with a 
long lifetime appears in the continuum.   

Figure 3 plots the average relative momentum as a function of 
$V_p$. As expected, when the resonance state exists both of the average 
relative momenta in the longitudinal and transverse directions are very 
small. It is important to note, 
however, that the transverse momentum remains finite even 
when the resonance energy is very close to zero and thus its 
lifetime becomes very long. This indicates that the resonant breakup 
mechanism is too much oversimplified. When the potential depth is deep 
enough to have 
a bound excited state, the average relative momentum shows quite different 
behaviour between the longitudinal and transverse directions. The 
longitudinal momentum difference is still very small and
changes continuously from the case where the excited state is the resonance.
In contrast with this the transverse momentum difference increases 
discontinuously from the resonance case. These results of small  
longitudinal and large transverse relative momenta might suggest that the 
breakup proceeds through the free-particle breakup mechanism in this case. 
There is, however, no physical reason that the free-particle breakup 
mechanism is correct. This is because two breakup processes occur when 
there is a bound excited state: One is the direct breakup to the continuum. 
The other is the breakup via the bound excited state. The latter process 
apparently does not fit in with the free-particle breakup mechanism.   
At present we do not have a simple explanation for 
the discontinuous change in the transverse momentum difference and the 
continuous behaviour in the longitudinal momentum difference which show up 
when the excited $0p$ orbit crosses the threshold.
When the $0p$ orbit is bound more deeply ($V_p < -35$ MeV), or
is in the nonresonant continuum ($V_p > -25$ MeV), the
momentum differences become similar to those of the previous subsection 
($V_p=V_0=-58.3$ MeV, see Fig. 1), that is, $\langle {k}_{\parallel}\rangle 
=0.026$ fm$^{-1}$ and $\langle {k}_{\perp}\rangle = 0.059$ fm$^{-1}$, and 
are rather insensitive to the potential depth. 

The $B(E1)$ strength function is compared in Fig. 4 for two choices of 
$V_p$: One is $V_p$=$-30.0$ MeV where the $0p$ resonance  appears at 
about 0.2 MeV. The other is $V_p = -32.0$ MeV which locates 
the $0p$ excited state at $E$=$-0.183$ MeV. The latter case corresponds to 
the previous calculation in \cite{Kido}. This was chosen because  the 
$\frac{1}{2}^{-}$  excited state of $^{11}$Be is known to have 
the strong $E1$ transition strength. 
The $B(E1)$ strength calculated with the $l$-independent potential 
is already presented in Fig.2.
When the resonance exists, the sharp peak appears at the resonance
energy. Except for the energy region of the sharp peak, the $B(E1)$  
strength is rather similar between the two cases of the  
resonance and the bound excited state. The $B(E1)$ strength function with 
the $l$-independent potential has also similar shape, although its 
magnitude is larger by a constant factor. 
This difference is understood by noting that, for the case of the 
$l$-independent 
potential, there is no resonant nor bound excited state in the $p$-wave 
so that all the $E1$ strength appears in the continuum.
The similarity in the shape of the $B(E1)$ strength for three cases 
indicates that the energy dependence of the non-resonant part of the 
$B(E1)$ strength is mainly determined by the wave function of the ground 
state, and is rather insensitive to the level structure of the excited 
states.

\bigskip
\begin{center}
{\bf IV. ANALYSIS OF $^{11}$Be COULOMB BREAKUP REACTION}
\end{center}
\quad

In this section we analyze the Coulomb breakup reaction of 
$^{11}$Be + $^{208}$ Pb  at the incident energy of 72 MeV/nucleon  
done at RIKEN \cite{Nakamura}. To reproduce the known properties
of the $^{11}$Be structure, we choose the potential between the neutron
and the $^{10}$Be core nucleus in the
following way: The depth $V_0$ is determined to reproduce the neutron 
separation energy of the $^{11}$Be (0.503 MeV) by assuming the $1s$ 
orbit of the 
halo neutron. For $l \ne 0$ channels the central and $ls$ potentials are 
included. The strength of the
$ls$ potential is set the standard value for the $p$-shell nucleus,
$V_{ls}=$32.8 MeV${\cdot}$fm$^{2}$. The strength of the central potential 
is then determined to reproduce the observed $\frac{1}{2}^{-}$ excited state 
located at $-$0.183 MeV from the neutron threshold. The state is assumed 
to be described simply with the $0p_{\frac{1}{2}}$ orbit. No deformation 
or clustering effect of the $^{10}$Be core nucleus is taken into account. 

The time evolution of the wave function is calculated by using 
the same parameter sets as the previous section. The 
Coulomb breakup cross section is obtained by the integration over the impact
parameter larger than $b_{\rm min} = 12$ fm (see Eq. (10)). The reaction 
of the impact parameters smaller than  
$b_{\rm min}$ proceeds by the nuclear force as well as the Coulomb force, 
and is assumed to lead to more violent nuclear reaction processes.

The Coulomb breakup process is dominated by the dipole component of the
target Coulomb field, and proceeds through the excitation to two angular 
momentum channels, $j^{\pi}=\frac{1}{2}^{-}$ and $\frac{3}{2}^{-}$. 
The $B(E1)$ transition strength to the bound level of $\frac{1}{2}^{-}$ is
calculated to be 0.251 $e^2$fm$^2$ in the present model, which is 
considerably larger than the measured value, 
0.115$\pm$0.011 $e^2$fm$^2$ \cite{Millener}. 
The transitions to the other bound levels below 
the ground state, which should not occur in principle, are not excluded in 
the process of the time evolution of the wave function, but the mixing-in  
of those states is found to be negligible in the present calculation.

We show in Fig.5 the impact parameter dependence of the breakup cross 
section $d\sigma_{\rm BU}/db=2{\pi}bP_{\rm BU}(b)$. The calculated 
distribution is compared with the data \cite{Nakamura}
which are extracted from the 
measured breakup cross sections by using the classical argument for the 
trajectory. Figure 6 compares the energy spectrum of the breakup cross 
section, $ d\sigma_{\rm BU}/dE $, with the measurement \cite{Nakamura}. 
The calculation reasonably reproduces the measured distribution. 
Dashed curves in Figs. 5 and 6 show the previous results of \cite{Kido} 
obtained without use of the $ls$ potential, where the $0p$ orbit was fitted 
to the energy of the  bound excited $\frac{1}{2}^{-}$ level. Since the 
$E1$ transition to this excited state, though fairly strong, does not 
lead to the breakup 
reaction, the breakup probability was underestimated in the previous 
calculation. By the introduction of the $ls$ potential the $0p$ orbit now 
splits into two levels, $0p_{\frac{1}{2}}$ and $0p_{\frac{3}{2}}$, in the 
present calculation. Since the bound  
$0p_{\frac{3}{2}}$ orbit is located below the ground state and the 
transition to this state is small,   
most of the $E1$ strength in the $p_{\frac{3}{2}}$
channel is distributed in the continuum. This is the reason why the breakup 
cross section increased in the present model 
which includes the $ls$ potential. The calculated $B(E1)$ strength 
function shown in Fig. 7 also confirms that the inclusion of the $ls$ 
potential leads to the increase in the strength for the same reason 
mentioned above. The $B(E1)$ strength is compared to the experimental 
data which are extracted from the breakup cross sections by using the 
method of virtual photon spectra \cite{Jackson}. The agreement between 
theory and 
experiment is rather good, which is expected because the energy spectrum of 
the breakup cross section has already shown reasonable agreement as shown 
in Fig. 6. 
We note, however, that the present model overestimates the $B(E1)$ 
strength to the $\frac{1}{2}^-$ excited state, and do not know what effects 
the overestimation causes on the breakup mechanism. 

We display in Fig. 8 the longitudinal and the transverse momentum 
distributions of the neutron-core relative motion. The impact parameter
is set the smallest value, $b=12$ fm. The longitudinal momentum distribution 
is obtained by integrating the distribution of Eq. (9) with respect to $k_x$ 
and $k_y$, while the transverse distribution is obtained by the integration 
over $k_y$ and $k_z$. The longitudinal momentum distribution slightly 
shifts to the negative direction, whereas the transverse momentum 
distribution shifts to the positive direction. The average value of the 
momentum of the relative motion is $\langle k_{\parallel}\rangle = -0.019$ 
fm$^{-1}$, and $\langle k_{\perp}\rangle = 0.053$ fm$^{-1}$, respectively. 
The measured difference of 
the longitudinal momentum is about 0.04 fm$^{-1}$, which is close to 
the value of  
$k_c = 0.046$ fm$^{-1}$ expected from the classical picture. Though the 
calculation reproduces a right order of magnitude, it explains only a half 
of the measured value.

Finally we consider the breakup mechanism referring to the result 
of the previous section. The mechanism is different in two channels, 
$j^{\pi}={\frac{1}{2}}^{-}$ and ${\frac{3}{2}}^{-}$, because they 
have different level 
structure of the excited states. In the $j^{\pi}=\frac{1}{2}^{-}$ channel 
there is a bound excited state which is very close to the threshold. This  
situation is similar to the case of $V_p = -32.0$ MeV which we already 
investigated in Fig. 3 with the $ls$ potential turned off. As expected, 
the momentum difference in this channel, especially in the longitudinal 
direction, is very small. On the other hand, there is no bound excited state 
in the $j^{\pi}=\frac{3}{2}^{-}$ channel. The situation is thus similar 
to the case of $V_p = -45$ MeV in Fig. 3. The average longitudinal 
momentum difference increases and turns out to be $-0.022$ fm$^{-1}$, about 
a half of the classical value of $k_c$. The breakup reaction of 
$^{11}$Be is thus considered to occur in these two different mechanisms. 
Our previous treatment \cite{Kido} which did not include the $ls$ potential 
is equivalent physically to including only the former process. It is 
therefore understandable that we obtained the very small longitudinal 
momentum difference in that case. 
A realistic choice of the internal Hamiltonian, 
particularly the inclusion of the spin-orbit interaction, is very important 
for a quantitative analysis of the breakup reaction of the $^{11}$Be nucleus.

\bigskip
\begin{center}
{\bf V. SUMMARY}
\end{center}
\quad

We investigated the Coulomb breakup mechanism of the nuclei with
single neutron-halo structure, focusing on the mechanism which causes
the momentum difference of the neutron-core relative motion after the 
breakup. The breakup process was described in the framework of the 
time-dependent Schr\"odinger equation by treating the target Coulomb 
field as a time-dependent external potential.

We first discussed the validity and the limitation of the classical
arguments made for the post-acceleration effect from the quantum-mechanical 
viewpoint by investigating the dependence of the reaction mechanism on 
the neutron binding energy and the level structure of the halo nucleus. 
We found that the classical arguments do not have quantitative accuracy 
though it is useful to understand the qualitative aspects of the
quantum calculation.  A behaviour which cannot be understood
in the classical picture occurred particularly when the bound excited state 
locates close to the neutron threshold. Small longitudinal and large 
transverse momentum differences were found in this case.

We analyzed the Coulomb breakup reaction of $^{11}$Be + $^{208}$Pb
by employing a realistic potential that describes the relative motion 
between the neutron and the $^{10}$Be nucleus. A spin-orbit potential 
was included to describe the splitting of the structure of $p$ states.    
The reaction proceeds through two channels, $j^{\pi}=\frac{1}{2}^{-}$ and 
$\frac{3}{2}^{-}$.
Since the $\frac{1}{2}^{-}$ channel has a bound excited state near the 
threshold but the $\frac{3}{2}^{-}$ channel does not have any bound 
excited states or low-lying resonances, the
breakup mechanism is rather different in the two channels.
By including the neutron-$^{10}$Be $ls$ potential we took into 
account, in the present analysis, the difference in 
the breakup mechanism which is sensitive to the level structure of the 
nucleus. By this improvement the post-acceleration effect was enhanced 
compared to the previous case which neglected the $ls$ potential and 
moreover the magnitude of the energy spectrum of the fragments  
was in reasonable agreement with the measured values. The momentum
difference in the longitudinal direction was, however, still 
underestimated by a factor of two compared with experiment.

\newpage
\begin{center}
{\bf Table I}
\end{center}

The classical estimates of the average longitudinal and transverse momenta, 
$\langle {k}_{\parallel}\rangle$ and $\langle {k}_{\perp}\rangle$, of the 
relative motion after the projectile nucleus fragments into 
the neutron and the core nucleus by the target Coulomb field. 
Three reaction mechanisms are classified and characterized by 
the classical momentum $k_c = [m_n/(m_n+M_C)]Z_T 
Z_C e^2/\hbar b v$, where $Z_Ce$ $(Z_Te)$ is the charge of the core (target) 
nucleus, $m_n$ and $M_C$ are the masses of the neutron and the core 
nucleus, respectively, 
$v$ is the incident velocity of the projectile nucleus, and $b$ is the 
impact parameter. The zero momentum for the resonant breakup is the 
limiting case of a long lifetime.  
\begin{center}
\begin{tabular}{cccc} \hline
              &\ \   Resonant  &\ \   Direct  & \ \  Free-particle  \\ \hline \hline
$\langle {k}_{\parallel}\rangle$& 0   & $k_c$   & 0     \\ \hline
$\langle {k}_{\perp}\rangle$    & 0   & $k_c$   & 2$k_c$\\ \hline
\end{tabular}
\end{center}
\vfill\eject

\newpage
\begin{center}
Figure Captions
\end{center}

{Fig.1.} The average values of the longitudinal (solid curve) 
and the transverse (dashed curve) momentum difference between the 
neutron and the core nucleus as a function of the neutron-core potential 
depth $V_0$.  The impact parameter $b$ is set 12 fm. 
The energies of the bound orbits from the 
neutron threshold are shown in the lower part.

{Fig.2.} The $B(E1)$ strength as a function of the energy $E$ of the  
relative motion between the neutron and the core nucleus. Two choices of   
the neutron-core potential depth $V_0$ are made to locate the $1s$ ground 
state at $-0.503$ MeV ($V_0=-58.3$ MeV) and $-2.00$ MeV 
($-65.2$ MeV).

{Fig.3.} The average values of the longitudinal (solid curve)
and the transverse (dashed curve) momentum difference between the neutron
and the core nucleus as a function of the neutron-core potential 
depth $V_p$ of $l \ne 0$ channel. The impact parameter $b$ is set 12 fm. 
The solid curve in the lower part shows the energy of the bound $0p$ 
orbit from the neutron threshold. 
The energy of the $1s$ orbit is fixed to $-0.503$ MeV as indicated 
by dashed line in the lower part.

{Fig.4.} The $B(E1)$ strength as a function of the energy $E$ of the 
relative motion between the neutron and the core nucleus. Two choices 
of the neutron-core potential depth $V_p$ of $l \ne 0$ channel are made to 
locate the $0p$ orbit at $-0.183$ MeV ($V_0=-32.0$ MeV) and $0.2$ MeV 
($-30.0$ MeV). The dotted 
curve indicates the strength reduced to one-tenth of the result shown by 
dashed curve.

{Fig.5.} The Coulomb breakup cross section as a function of the
impact parameter $b$.  
The solid curve is the result of the present model, while 
the dashed curve is the result with the $ls$-potential turned off \cite{Kido}. 
Experimental data are from \cite{Nakamura}.

{Fig.6.} The Coulomb breakup cross section as a function of the
energy $E$ of the relative motion between the neutron and the $^{10}$Be 
nucleus. The solid curve is the result of the present model, while 
the dashed curve is the result with the $ls$-potential turned off \cite{Kido}. 
Experimental data are from \cite{Nakamura}.

{Fig.7.} The $B(E1)$ strength as a function of the energy $E$ of the 
relative motion between the neutron and the core nucleus. Experimental 
data are from \cite{Nakamura}.

{Fig.8.} The longitudinal (a) and transverse (b) momentum distributions 
of the relative motion between the neutron and the $^{10}$Be nucleus. 
The impact parameter $b$ is set 12 fm. 
The solid curve is the result of the present model, while 
the dashed curve is the result with the $ls$-potential turned off \cite{Kido}.

\end{document}